\documentclass[prl,twocolumn,aps,longbibliography,superscriptaddress,notoc]{revtex4-2} % for arxiv

\usepackage{bm}
\usepackage{graphicx}
\usepackage{amssymb}
\usepackage{amsmath}
\usepackage{eufrak}
\usepackage{color}
\usepackage[utf8]{inputenc}
\usepackage[unicode=true,colorlinks=true,citecolor=blue,urlcolor=blue]{hyperref}

\let\ifr\i
\renewcommand{\i}{{\rm i}}
\renewcommand{\d}{\mathrm d}
\renewcommand{\emph}{\textit}
\usepackage{braket}

\usepackage{chngcntr}
\newcommand{\enquote}{}
\graphicspath{{figs/}}

\newcommand{\nix}[1]{}

\begin{document}

\title{Evidencing the squeezed dark nuclear spin state in lead halide perovskites}

\author{E. Kirstein}
\affiliation{Experimentelle Physik 2, Technische Universit\"at Dortmund, 44221 Dortmund, Germany}

\author{D. S. Smirnov}
\email[Electronic address: ]{smirnov@mail.ioffe.ru}
\affiliation{Ioffe Institute, Russian Academy of Sciences, 194021 St. Petersburg, Russia}

\author{E. A. Zhukov}
\affiliation{Experimentelle Physik 2, Technische Universit\"at Dortmund, 44221 Dortmund, Germany}
\affiliation{Ioffe Institute, Russian Academy of Sciences, 194021 St. Petersburg, Russia}

\author{D. R. Yakovlev}
\affiliation{Experimentelle Physik 2, Technische Universit\"at Dortmund, 44221 Dortmund, Germany}
\affiliation{Ioffe Institute, Russian Academy of Sciences, 194021 St. Petersburg, Russia}
\email{dmitri.yakovlev@tu-dortmund.de}

% \author{M. M. Glazov}
% \affiliation{Ioffe Institute, Russian Academy of Sciences, 194021 St. Petersburg, Russia}

\author{N. E. Kopteva}
\affiliation{Experimentelle Physik 2, Technische Universit\"at Dortmund, 44221 Dortmund, Germany}

\author{D. N. Dirin}
\affiliation{Department of Chemistry and Applied Biosciences,
Laboratory of Inorganic Chemistry, ETH Z\"{u}rich, 8093 Z\"{u}rich, Switzerland}

\author{O. Hordiichuk}
\affiliation{Department of Chemistry and Applied Biosciences,
Laboratory of Inorganic Chemistry, ETH Z\"{u}rich, 8093 Z\"{u}rich, Switzerland}

\author{M. V. Kovalenko}
\affiliation{Department of Chemistry and Applied Biosciences,
Laboratory of Inorganic Chemistry, ETH Z\"{u}rich, 8093 Z\"{u}rich, Switzerland} 
\affiliation{Department of Advanced Materials and
Surfaces, Laboratory for Thin Films and Photovoltaics, Empa - Swiss Federal Laboratories for Materials Science and Technology, 8600 D\"{u}bendorf, Switzerland}

\author{M. Bayer}
\affiliation{Experimentelle Physik 2, Technische Universit\"at Dortmund, 44221 Dortmund, Germany}
\affiliation{Ioffe Institute, Russian Academy of Sciences, 194021 St. Petersburg, Russia}

\begin{abstract}
Coherent many-body states are highly promising for robust and scalable quantum information processing. While far-reaching theoretical predictions have been made for various implementations, direct experimental evidence of their appealing properties can be challenging. Here, we demonstrate coherent optical manipulation of the nuclear spin ensemble in the lead halide perovskite semiconductor FAPbBr$_3$ (FA=formamidinium), targeting a long-postulated collective dark state that is insensitive to optical pumping. Via optical orientation of localized hole spins we drive the nuclear many-body system into an entangled state, requiring a weak magnetic field of only a few Millitesla strength at cryogenic temperatures. During its fast build-up, the nuclear polarization along the optical axis remains small, while the transverse nuclear spin fluctuations are strongly reduced, corresponding to spin squeezing as evidenced by a strong violation of the generalized nuclear squeezing-inequality with $\xi_s < 0.3$. The dark state evidenced in this process corresponds to an approximately 750-body entanglement between the nuclei. Dark nuclear spin states can be exploited to store quantum information benefiting from their long-lived many-body coherence and to perform quantum measurements with a precision beyond the standard limit.
\end{abstract}

\maketitle

\section{Introduction}

The range of many-body states in condensed matter is formidably rich: the fractional quantum Hall~\cite{Stormer1999,Hansson2017} and the Wigner crystal~\cite{Wigner1934,Shapir2019,RubioVerdu2021} states are just two examples~\cite{Abanin2019}. Quantum coherence is essential for the formation of many of them.

The many-body state of localized spins in solid state is arguably one of the best controlled and potentially scalable hardware to that end. For example, an electron spin can be used to create entanglement between multiple photons for one-way quantum computing~\cite{Lindner2009,Schwartz2016,Raussendorf2001}. However, electron and photon spins suffer from fast decoherence or short lifetime, limiting their applicability for quantum information storage and many-body unitary operations. Nuclear spins, on the other hand, are largely isolated from their environment resulting in extended coherence times. Thus, they complement electron spins~\cite{PhysRevLett.91.246802,Witzel2007,Wang2012}, particularly due to possible electron-nuclear spin interfacing~\cite{book_Glazov,Urbaszek,Chekhovich_protocol,Gangloff62,chekhovich2020nuclear}.

Spin-based concepts for solid state quantum computing and quantum metrology rely on many-body entanglement in combination with nuclear spin squeezing. The latter of which is the ultimate goal of nuclear spin manipulation~\cite{GUHNE20091,MA201189}. Almost 20 years ago, a coherent many-body dark nuclear spin state (DNSS) was predicted that can be formed by orienting optically the spins of resident charge carriers interacting with the nuclear spin bath in the host lattice~\cite{PhysRevLett.91.017402,PhysRevLett.91.246802}.
Mathematically, the DNSS is the state which vanishes upon action of the rising operator of the total nuclear spin. That is implemented by the optical pumping in the experiment: DNSS formation blocks the optical nuclear spin pumping in analogy with optically dark states in ensembles of atoms, which are immune to the light-matter interaction~\cite{Mandel1995}.
The DNSS belongs to the class of maximally entangled singlet states, so that it can be exploited not only to suppress the dephasing of localized electron and hole spin qubits~\cite{Kurucz2009,Yu2013}, but also for storage of quantum information~\cite{PhysRevLett.91.246802,Witzel2007} or for quantum metrology applications~\cite{Ding2020}.

\begin{figure*}[t]
  \includegraphics[width=\linewidth]{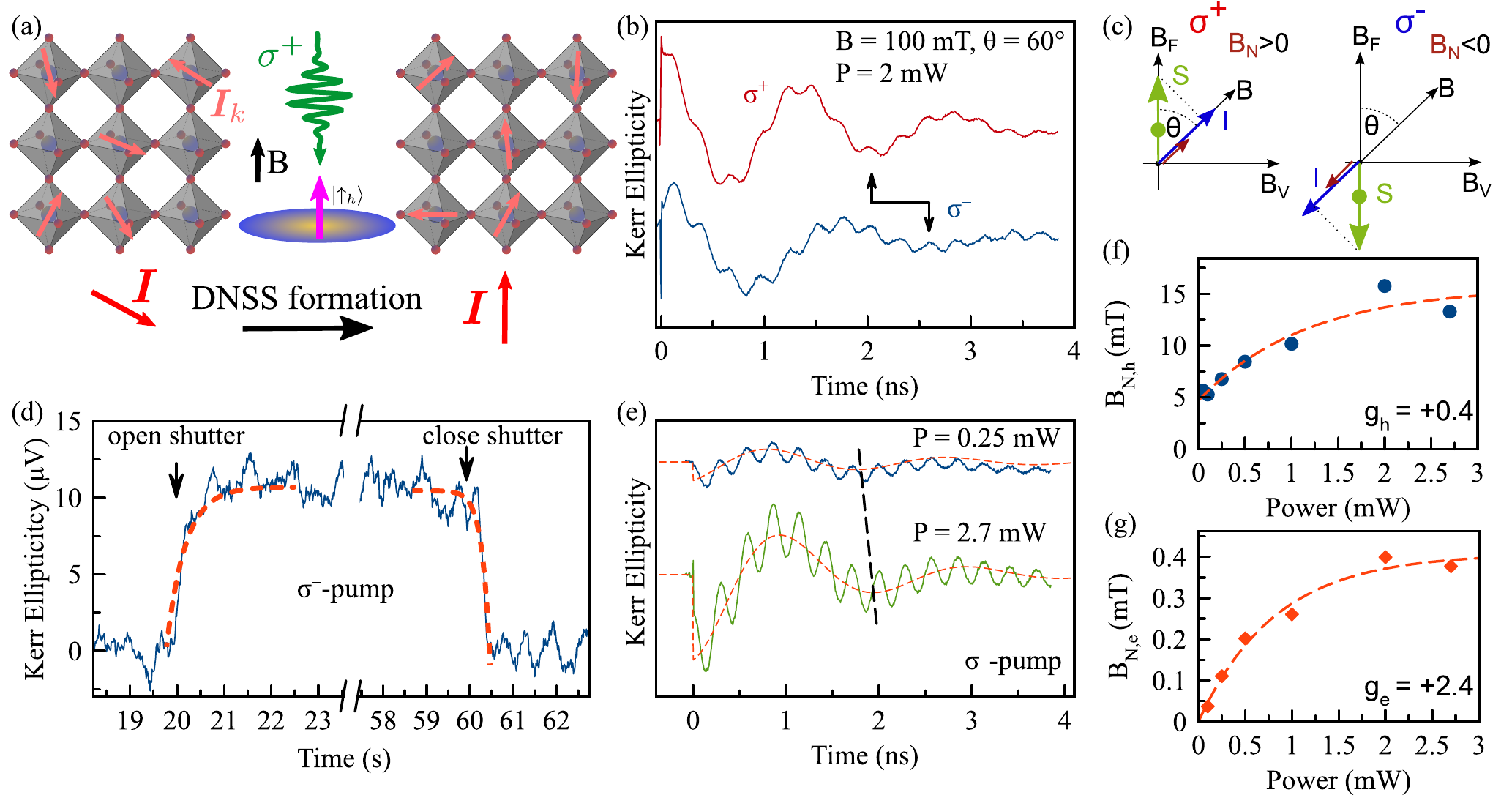}%\\
  \caption{\textbf{Measurement of the nuclear spin polarization in FAPbBr$_3$.} (a) Formation of the DNSS: after optical orientation (green light pulse) of a localized hole (magenta arrow), individual nuclear spins $\bm I_k$ (small coral arrows) rotate through the hyperfine interaction, each by the same angle, so that the total nuclear spin polarization $\bm I$ (large red arrow) is directed along the optical axis. (b) Time-resolved Kerr ellipticity dynamics measured for $\sigma^+$ and $\sigma^-$ pump pulse polarization measured at 2.191~eV photon energy, in which oscillations associated with electron and hole spin precession about the magnetic field are superimposed. Arrows mark the second oscillation minimum in the hole signal after the pump pulse at time zero. (c) Illustration of nuclear spin polarization in tilted magnetic field for $\sigma^+$ (left) and $\sigma^-$ (right) pump polarization. (d) Time-resolved, measurement of the Overhauser field rise and decay, showing changes on time scales below the 2\,ms low pass of the lock-in amplifier. Experimental parameters: $T=1.6$\,K, $P=2$\,mW, $B=0.1$\,T, $\theta=60^\circ$, $\Delta t=3.6$\,ns. (e) Time-resolved Kerr ellipticity dynamics measured at different pump powers for fixed pump polarization. Dashed line indicates the temporal shift of the hole oscillation for increased pump power. (f,g) Overhauser field for holes and electrons, respectively, extracted from the Kerr ellipticity dynamics, using the hole and electron $g$-factors of $g_h=+0.4$ and $g_e=+2.4$.}
\label{fig:dnptilted}
\end{figure*}

The DNSS is characterized by destructive interference of the nuclear spin amplitudes in the collective transverse components, which leads to their strong suppression without large nuclear spin polarization [Fig.~\ref{fig:dnptilted}(a)]. The destructive interference arises from quantum correlations and entanglement between the nuclear spins, somewhat analogous to the spin structure in antiferromagnetic materials. Realization and demonstration of the DNSS have turned out to be challenging due to threats from the dipole-dipole and quadrupole nuclear interactions. In prior works on III-V semiconductor quantum dots~\cite{SkibaSzymanska2008,PhysRevLett.104.066804,Hildmann2014}, where the nuclear spin polarization was measured, the limited polarization value was attributed to DNSS formation, but without proof of suppression of the transverse spin fluctuations~\cite{PhysRevLett.91.017402}.
%it was suggested that the DNSS blocks a large nuclear spin polarization,
Later, detailed investigations of GaAs quantum dots revealed 80\% nuclear polarization without any signatures of DNSS formation, i.e. the nuclear spin state in this system could be described by an effective nuclear spin temperature and the absence of correlations~\cite{Chekhovich_constants}. Correlations between nuclear spins was evidenced for non-thermal nuclear spin states~\cite{maletinsky2009breakdown}, narrowed nuclear spin states~\cite{Reilly817,xu09,Togan2011}, and nuclear frequency focusing~\cite{A.Greilich09282007}. Thus none of these studies evidenced the DNSS. In a recent study~\cite{Gangloff2021}, the authors reconstructed the nuclear spin statistics and suggested destructive interference between different nuclear spins. However, a measurement of the transverse spin components, mandatory for DNSS demonstration, is missing.

Here we study the intertwined spin dynamics of nuclei and charge carriers in a FAPbBr$_3$ lead halide perovskite crystal, using tailored optical pump-probe schemes. More specifically, we develop a method for assessing the nuclear spin inertia, which allows us to access the regime of weak magnetic fields in the milliTesla range. We find that excitation with circularly polarized light leads to the disappearance of the transverse nuclear spin fluctuations with simultaneous absence of a significant longitudinal nuclear spin polarization. The combination of these two factors directly evidences the destructive interference of nuclear spins and thus the DNSS formation. This many-body correlated state builds up fast on short time scales in the ms-range and shows significant spin squeezing evaluated through the Kitagawa-Ueda parameter~\cite{PhysRevA.47.5138} $\xi_s=0.29$, corresponding to entanglement of about $750$ nuclear spins.

\section{Results}

The exciton resonance in the hybrid organic-inorganic lead halide perovskite FAPbBr$_3$ crystals is located at 2.191\,eV. We have measured the time-resolved Kerr ellipticity (TRKE) at this energy. We give basic information on the charge carrier spin dynamics and the photoluminescence spectrum in the Supplementary Information I.A. The nuclear spin polarization has been addressed by the Overhauser field experienced by the resident hole spins, which we observe in TRKE, recorded at the temperature of $T=1.6$\,K in tilted magnetic field, Figure~\ref{fig:dnptilted}(b). The spin dynamics consists of slow and fast oscillating components related to the hole (h) and electron (e) spin precession about the total magnetic field. The corresponding carrier spin $g$-factors are $g_h=+0.4$ and $g_e=+2.4$ (Supplementary Information I.A).

%The low-temperature photoluminescence spectrum has a maximum at around 2.18\,eV, see the Supplementary Information I.A, where also basic information on the charge carrier spin dynamics is given. 

\begin{figure*}[t]
	\includegraphics[width=\linewidth]{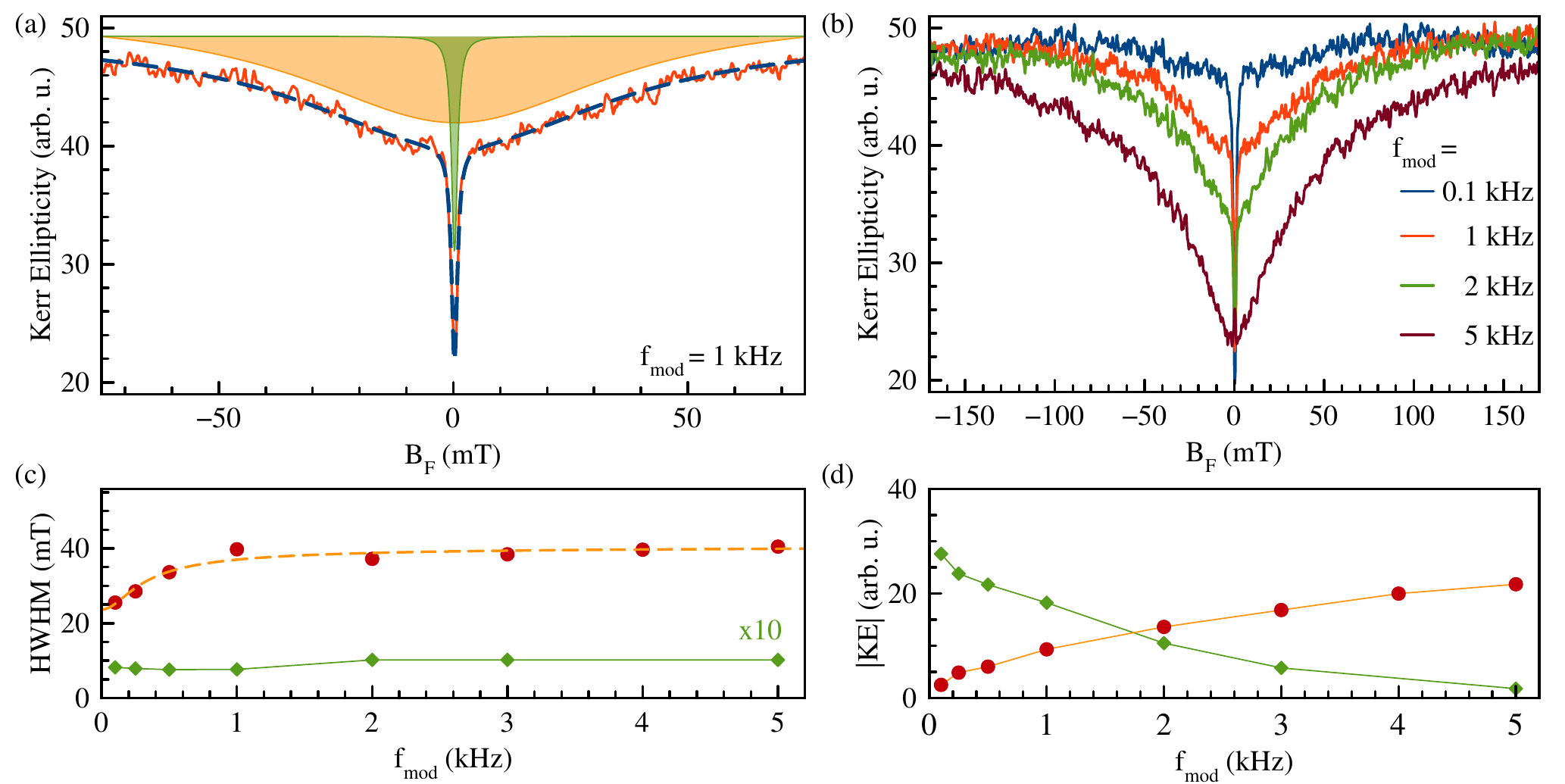}
\caption{\textbf{Nuclear spin inertia measurement.} (a) Representative PRCs each consisting of a broad (orange filling) and a narrow (green filling) component. (b) PRCs at different polarization modulation frequencies. (c,d) Polarization modulation frequency dependencies of widths and amplitudes of the broad (red symbols) and narrow PRC components (green symbols). Lines are guides to the eye. $T=5$\,K.}
\label{fig:nucin}
\end{figure*}

The strong spin-orbit interaction in perovskites provides a selectivity for light absorption of different circular polarizations~\cite{Nestoklon2018}. As a result, optical spin orientation is feasible for resident electrons and holes, localized at different sites in the sample at low temperatures, which  represents a situation similar to that provided by quantum confinement. The angular momentum of a $\sigma^+$ or $\sigma^-$ photon is transferred to the charge carrier spin polarization, and further to the host lattice nuclei through the hyperfine interaction, which leads to build-up of an Overhauser field $\bm B_N$ parallel or antiparallel to the external magnetic field, see Fig.~\ref{fig:dnptilted}(c). This field can be directly determined through measurement of the electron and hole Larmor precession frequencies about the total magnetic field, to which the external field and the Overhauser field contribute. With increasing pump power, the Overhauser field rises [Fig.~\ref{fig:dnptilted}(e)], but saturates at relatively small values of $B_{N,h}=15$\,mT for the holes [Fig.~\ref{fig:dnptilted}(f)] and $B_{N,e}=0.4$\,mT for the electrons [Fig.~\ref{fig:dnptilted}(g)]. Similar to other perovskites~\cite{belykh2019coherent,Aebli2020}, the hyperfine interaction is much stronger for the holes due to the dominant s-type orbital contribution to the Bloch wave functions at the top of the valence band. Thus, we focus on the holes in our study, which dominate the Kerr ellipticity dynamics. The measured dependence of the Overhauser field on the direction and strength of the external field is presented in the Supplementary Information [Fig.~\ref{fig:AngleDep}].

The small measured Overhauser field, as compared to the maximum possible strength of 1.3\,T evaluated from the hyperfine coupling constants~\cite{kirstein2022}, indicates an unusual nuclear spin statistics which can be highlighted further by measurement of its rise and decay times. Therefor, we first close the pump beam to erase the nuclear spin polarization, and then open the shutter after which we observe almost immediately a nuclear spin polarization, see Fig.~\ref{fig:dnptilted}(d). After closing the shutter again, the spin polarization decays on the same time scale of about $100$\,ms. This nuclear spin response time is much shorter than the previously reported longitudinal nuclear spin relaxation times of $T_{1,N}=5$~s in FA$_{0.9}$Cs$_{0.1}$PbI$_{2.8}$Br$_{0.2}$~\cite{kirstein2022} and $960$\,s in MAPbI$_3$~\cite{Kirstein2022ACS}. In fact, the variation is limited by the shutter mechanical opening/closing times in experiment, so that the nuclear spin dynamics occur even faster.

To access this surprisingly fast nuclear spin response time, we develop a technique for measuring the nuclear spin inertia. Similarly to the electron spin inertia technique~\cite{Korenev2015,PhysRevB.98.121304}, it monitors the Kerr ellipticity signal as function of the longitudinal magnetic field (Faraday geometry) with simultaneous modulation of the pump helicity at different frequencies $f_{\rm mod}$. An example for $f_{\rm mod}=1$\,kHz is shown in Fig.~\ref{fig:nucin}(a). The application of a longitudinal magnetic field leads to a spin polarization recovery due to the suppression of nuclei-induced spin relaxation of the holes~\cite{PRC}. Strikingly, the shape of this polarization recovery curve (PRC) changes with the modulation frequency [Fig.~\ref{fig:nucin}(b)]: at the high frequency of $f_{\rm mod}=5$\,kHz, the PRC resembles a wide dip of Lorentzian shape with the half width at half maximum (HWHM) of 40\,mT. With decreasing frequency down to 0.1~kHz, this dip gradually disappears and is replaced by a much narrower dip, having the HWHM of about 1\,mT. Generally, a PRC  consists of the broad and the narrow component, each of which can be phenomenologically described by a Lorentzian. Their widths depend weakly on $f_{\text{mod}}$ [Fig.~\ref{fig:nucin}(c)], and their amplitudes change in opposite ways [Fig.~\ref{fig:nucin}(d)].

The spin relaxation time of a hole is of the order of $1\,\mu$s (as measured using the same spin inertia method at higher modulation frequencies), so that we attribute the PRC changes with modulation frequency to fast hole spin-induced nuclear spin dynamics occurring on time scales of 1\,ms, which is shorter than the typical quantum decoherence time of nuclear spins. Further details will be given in the Theory section, additional pump-power dependencies and measurements for $f_{\rm mod}=0$ are presented in the Supplementary Information [Figs.~\ref{fig:PRCconst} and~\ref{fig:PowerDep}].

\begin{figure}[t]
  \includegraphics[width=1\linewidth]{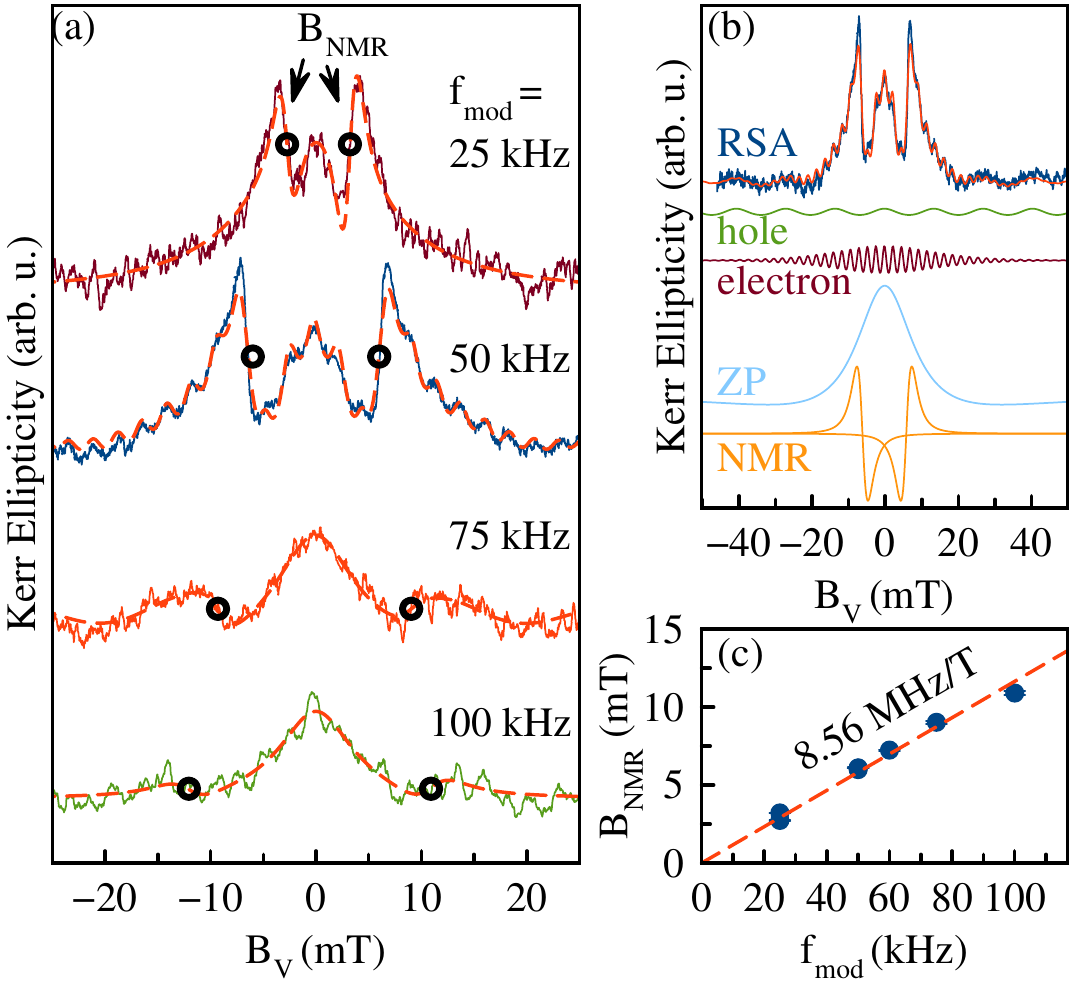}%{RSA_rotframe3}
\caption{\textbf{ODNMR measurement.} (a) Resonant spin amplification signals at different modulation frequencies. (b) Decomposition of the RSA signal at $f_{\rm mod}=50$\,kHz into four components by fitting. From top to bottom: RSA signal (blue) together with the sum of the four components (red), contributions of the hole (green), the electron (dark red), the zero peak [ZP] (light blue), the nuclear magnetic resonances [NMR] (orange). (c) Dependence of the resonance field on the modulation frequency, in agreement with the $^{207}$Pb gyromagnetic ratio. $T=5$\,K.}
\label{fig:RSA}
\end{figure}

Further, we identify the specific nuclear isotope dominating the hyperfine interaction. For this, we use the Voigt geometry with the magnetic field normal to the optical axis, where a nuclear magnetic resonance appears if the carrier polarization modulation matches the nuclear Zeeman splitting. The carrier polarization modulation equals the pump modulation $f_\textrm{mod}$~\cite{Heisterkamp2015}. The Kerr ellipticity signal [Fig.~\ref{fig:RSA}(a)] can be separated into an electron-related component and a hole-related component, both representing damped oscillations, in addition a nuclei-related zero field peak and nuclear magnetic resonance (NMR) appears [Fig.~\ref{fig:RSA}(b)]. The latter dominate the signal, their position is given by $B_{NMR}=f_{\text{mod}}/\gamma$ with $\gamma=8.56$~MHz/T [Fig.~\ref{fig:RSA}(a,c)]. This value closely corresponds to the gyromagnetic ratio of $^{207}$Pb given by 8.88\,MHz [see Supplementary Information, Tab.~\ref{tab:nmr}]. As in previous measurements of optically detected nuclear magnetic resonance in perovskites, the small chemical shift may occur due to the presence of a spin-polarized hole~\cite{kirstein2022}. 

When the light polarization modulation frequency is high, the nuclear spin distribution remains in equilibrium. The polarization recovery effect in Fig.~\ref{fig:nucin}(b) at $f_{\text{mod}}=5$\,kHz then demonstrates nuclei-dominated hole spin relaxation~\cite{PRC}: In the absence of the external magnetic field, the spin precession in the randomly oriented Overhauser field $\bm B_N$ leads to spin dephasing [Fig.~\ref{fig:teor}(a)], while application of a longitudinal field $\bm B$ rotates the total magnetic field experienced by the holes, $\bm B_{\text{tot}}=\bm B_N+\bm B$, towards the optical axis and suppresses the spin dephasing. Thus, the HWHM of the PRC in this case gives the typical fluctuation of the Overhauser field, $\Delta_B=40$\,mT (Supplementary Information II.A).

The PRC directly measures the average angle $\theta$ between the optical $z$ axis and the hole spin precession frequency vector, which includes the Overhauser and external field contributions [inset in Fig.~\ref{fig:teor}(f)]. The nuclear spin inertia measurements reveal the recovery of the hole spin polarization at magnetic fields above $1$\,mT, when decreasing the polarization modulation frequency from 5\,kHz down to 0.1\,kHz. This appears to be a paradox, since the typical nuclear spin polarization rate is much smaller than these rates. Moreover, the largest nuclear spin polarization of $B_{N,h}=15$\,mT shown in Fig.~\ref{fig:dnptilted}(e) is considerably smaller than the typical fluctuation of the Overhauser field $\Delta_B=40$\,mT. However, the suppression of the transverse Overhauser field fluctuations in absence of a significant nuclear spin polarization unambiguously evidences the destructive interference between nuclear spins due to their quantum correlations, and exactly matches the predictions for the DNSS formation~\cite{PhysRevLett.91.017402}.

\section{Theory}

The established concept of dynamic nuclear spin polarization relies on the assumption of an effective nuclear spin temperature and the absence of correlations between the nuclei. In the DNSS, by contrast, quantum correlations and destructive interference in the collective transverse nuclear spin components necessitate a purely quantum mechanical description. Thus, to model the coherent nuclear spin dynamics, we apply the central spin box model described by the following Hamiltonian:
\begin{equation}
  \label{eq:ham}
  \mathcal H=A \bm I \bm S+\hbar\Omega_{L,h} S_z.
\end{equation}
Here $\bm I$ is the total nuclear spin in the hole localization volume, $\bm S$ is the hole spin, $A$ is the hyperfine coupling constant of the holes, $\Omega_{L}$ is the hole Larmor precession frequency in the external magnetic field applied along the optical $z$ axis, and $\hbar$ is the reduced Planck constant. The spin precession about the nuclear field is included in the renormalization of $\Omega_{L}$. 

The total nuclear spin is formed by $N$ individual $^{207}$Pb $1/2$ spins $\bm I_k$: $\bm I=\sum_{k=1}^N\bm I_k$. The number of nuclear spins can be estimated from the PRC width using the hyperfine interaction constant $A=33\,\mu$eV~\cite{kirstein2022}: $N=[A/(g_h\mu_B\Delta_B)]^2=1300$. Using the lattice constant of $a_0=0.6$\,nm, this gives the hole localization length $l=a_0(N/0.22)^{1/3}=11$\,nm (with 0.22 being the natural lead spin abundance), which is similar to other perovskites~\cite{belykh2019coherent,kirstein2022}. Since $N\gg1$, Eq.~\eqref{eq:ham} implies hole spin precession between two pump pulses with the constant frequency $A\bm I/\hbar+\Omega_L\bm e_z$ ($\bm e_z$ is the unit vector along the $z$ axis), which leads to hole spin dephasing in weak fields for a randomly oriented total nuclear spin $\bm I$, close to equilibrium.

\begin{figure*}
  \includegraphics[width=\linewidth]{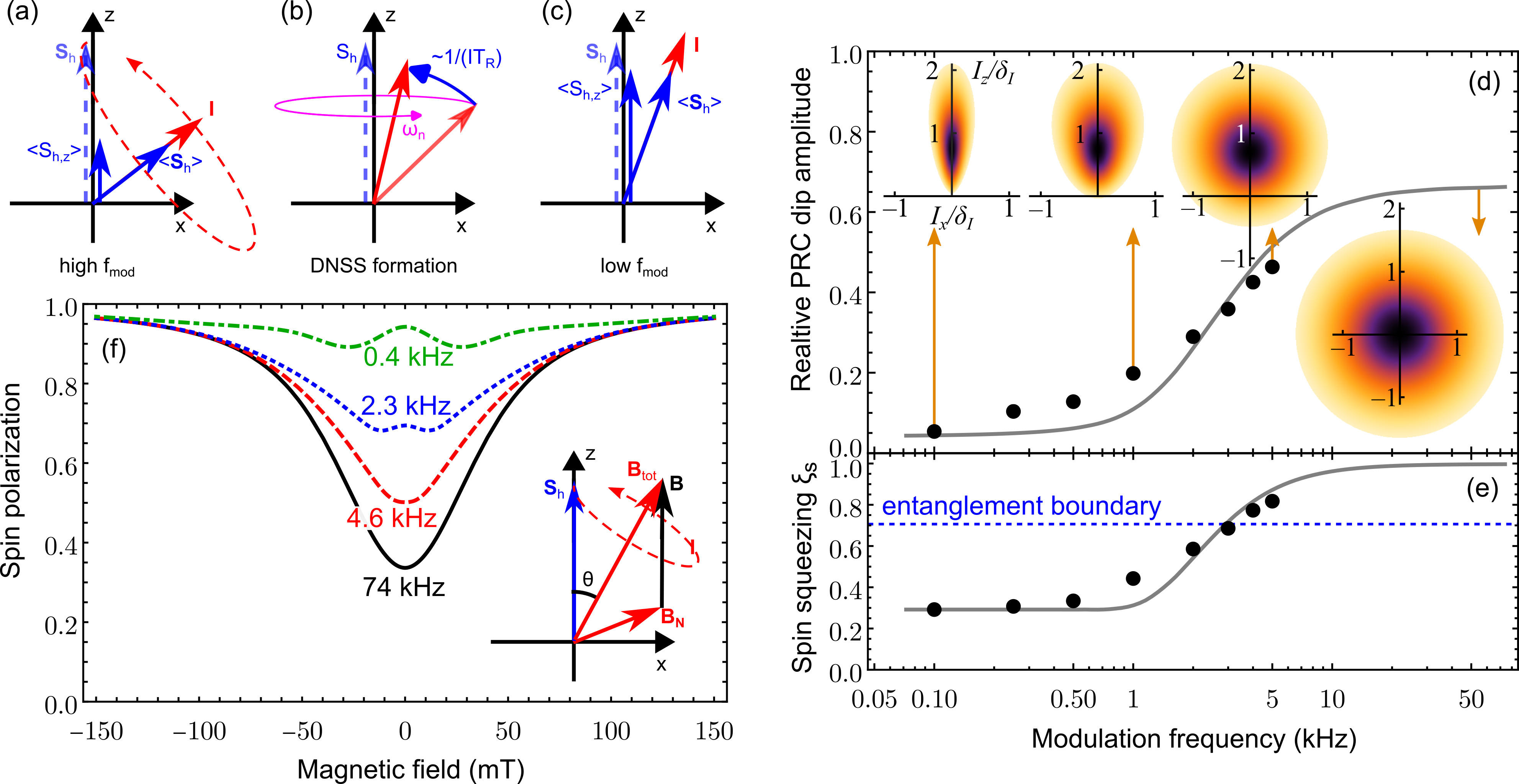}
  \caption{\label{fig:teor}
    \textbf{Modeling the nuclear spin inertia for DNSS formation.}
    (a) The hole spin precession in the randomly oriented Overhauser field leads to partial spin dephasing.
    (b) During the DNSS formation, the Overhauser field rotates towards the optical axis.
    (c) This leads to the suppression of hole spin dephasing.
    (d) Comparison of the measured (dots) and calculated (solid line) frequency dependencies of the amplitude of the broad PRC dip.
    The insets show the nuclear spin distribution functions at the corresponding modulation frequencies normalized to their maxima, evidencing the spin squeezing mostly in transverse direction.
    (e) Nuclear spin squeezing parameter $\xi_s$ calculated from the experimental data (dots) and calculated numerically (solid line). States below the blue dashed line are entangled.
    (f) Calculated frequency dependence of the PRC.
  }
\end{figure*}

Continuous hole spin pumping and its dephasing lead to the transfer of angular momentum to the nuclei, as can be seen from the conservation of the total angular momentum $I_z+S_z$ by the Hamiltonian~\eqref{eq:ham}. Simultaneously, the conservation of the absolute value of the total nuclear spin $I$ prevents the build-up of a nuclear spin polarization larger than $\sim1/\sqrt{N}$. Therefore, the total nuclear spin is rotated towards the $z$ axis, while its transverse components decrease, but its absolute value does not increase. The collective nuclear spin dynamics driven by the interaction with a single hole spin produces quantum correlations and leads to entanglement between the nuclei. As a result, the DNSS is formed [Fig.~\ref{fig:teor}(b)], in agreement with the original theoretical predictions~\cite{PhysRevLett.91.017402,PhysRevLett.91.246802}. 

The quantum mechanical solution of the box model~\cite{RS1983,Kozlov2007,yugova11,PhysRevLett.126.216804} allows us to fit the experimentally measured amplitude of the broad PRC dip as function of the polarization modulation frequency, see Fig.~\ref{fig:teor}(d). The modification of the nuclear spin distribution with decreasing modulation frequency is shown in the corresponding insets. The alignment of all nuclear spin fluctuations along the optical axis at low modulation frequencies cancels the hole spin precession [Fig.~\ref{fig:teor}(c)] and restores the hole spin polarization to the same value as in large magnetic fields, where the hole and nuclear spins become in effect decoupled (the hole spin dephasing by the nuclei is suppressed). Thereby also the whole PRC amplitude rises, as shown in Fig.~\ref{fig:teor}(f). From fitting the nuclear spin inertia we find the DNSS formation rate (typical frequency of nuclear spin rotation) of $\nu_0=0.9$\,ms$^{-1}$ (Supplementary Information II.C).

\section{Analysis and discussion}

The ratio of the hole spin polarization at a given magnetic field and in saturation (at $B\gtrsim 100$\,mT) represents a collective measurement of the total nuclear spin components $\braket{(I_x^2+I_y^2)/I^2}$~\cite{PRC}. This allows us to quantify the suppression of the transverse nuclear spin fluctuations by the Kitagawa and Ueda spin squeezing parameter $\xi_s^2=4\braket{I_x^2}/N$~\cite{PhysRevA.47.5138,MA201189} ($\braket{I_x^2}=\braket{I_y^2}$), which for uncorrelated spins equals to unity. This parameter extracted from the measured PRC amplitude is plotted in Fig.~\ref{fig:teor}(e) by the dots, and its simulated frequency dependence with the same DNSS formation rate is shown by the solid line (Supplementary Information II.D). For almost complete suppression of the broad PRC dip to values below $0.05$ at $f_{\rm mod}=100$\,Hz, the spin squeezing parameter is $0.29$, which is limited by nuclear spin diffusion due to incomplete hole spin polarization and quantum fluctuations of the transverse total nuclear spin components.

The quantum correlations between the nuclear spins suggest their entanglement. This can be shown using the generalized spin squeezing inequality~\cite{Toth2007,MA201189}
\begin{equation}
  \braket{I_x^2}+\braket{I_y^2}+\braket{(I_z-\braket{I_z})^2}\ge M/2.
  \label{eq:gss}
\end{equation}
Its violation requires $M$-body entanglement between $N$ nuclear spins, meaning that there are at least $M$ spins, which are entangled with the rest of the ensemble~\cite{Guehne2004,GUHNE20091}. To show its violation, we use the upper boundary for the longitudinal nuclear spin fluctuations, $\braket{(I_z-\braket{I_z})^2}\le N/4$, which yields the lower limit of $\xi_s^{(0)}=0.71$ for the entangled states. States with $\xi_s<\xi_s^{(0)}$ violate Eq.~\eqref{eq:gss} with $M=N$ and are entangled. The maximum achieved DNSS with $\xi_s=0.29$ is at least $M=750$-body entangled. However, our theoretical simulations of the nuclear spin fluctuations in the DNSS suggest much deeper entanglement.

There are good reasons for FAPbBr$_3$ perovskites to be the material system for experimental observation of the DNSS: (i) Lead has either nuclear spin $0$($^{206}$Pb, $^{208}$Pb with abundance of 77.9\%) or $1/2$ ($^{207}$Pb with abundance of 22.1\%), which excludes quadrupole splitting of the nuclear spin levels due to strain or electric field gradients. (ii) The abundance of nonzero lead spins is relatively low and the elementary cell is relatively large. This suppresses the nuclear dipole-dipole interactions, which threaten the DNSS. (iii) The magnetic fields that have to be applied are relatively weak of the order of the fluctuations of the Overhauser field, which accelerates the DNSS formation $\propto1/B^2$.

Our findings establish lead halide perovskites as a promising platform for exploration and exploitation of intertwined hole and nuclear spin dynamics to excite non-classical collective spin states with quantum correlations. The nuclear spin inertia method is powerful for clearly demonstrating DNSS formation, especially for application to other systems with $1/2$ nuclear spins including other perovskites or to unstrained quantum dots. In combination with existing demonstrations of nuclear spin based quantum registers~\cite{chekhovich2020nuclear,Ruskuc2022} and collective spin measurements~\cite{jackson2021quantum,Gangloff2021}, the DNSS may be the most reliable platform for quantum metrology, quantum information storage and processing with solid state spins beyond the standard quantum limit~\cite{Pezze2018}, due to the low decoherence of the DNSS, caused by destructive interference in the transverse spin fluctuations.

\renewcommand{\i}{\ifr}
\let\oldaddcontentsline\addcontentsline% Store \addcontentsline
\renewcommand{\addcontentsline}[3]{}% Make \addcontentsline a no-op

%\bibliography{Perovskite_FAPbBr_NMR}
%apsrev4-2.bst 2019-01-14 (MD) hand-edited version of apsrev4-1.bst
%Control: key (0)
%Control: author (8) initials jnrlst
%Control: editor formatted (1) identically to author
%Control: production of article title (0) allowed
%Control: page (0) single
%Control: year (1) truncated
%Control: production of eprint (0) enabled
%

\section{Methods}

\subsection{Growth of FAPbBr$_3$ crystals}

The FAPbBr$_3$, with FA being formamidinium, perovskite crystal was grown using the inverse temperature crystallization technique (sample code: OH0071a). In essence, the reactant salts (FABr and PbBr2) were dissolved in a mixture of DMF:GBL (1:1 v/v), forming the precursor solution. By rising the temperature of the solution, the sample crystallized due to retrograde solubility of the perovskite crystals in the chosen solvent mixture, see Ref.~\cite{saidaminov2015}. The studied crystal has a reddish color, see the inset in Fig.~\ref{fig:intro}(a), with a size of $5 \times 5 \times 2$~mm$^3$.

\subsection{Magneto-optical measurements} 

The sample was placed in a cryostat at a temperature variable from 1.6\,K up to 300\,K. For $T=1.6$\,K the sample was immersed in superfluid helium, while for temperatures in the range from  4.2\,K to 300\,K the sample was held in cooling helium gas. The cryostat is equipped with a vector magnet composed of three superconducting split coils orthogonal to each other. This allows us to apply magnetic fields up to 3\,T in any direction. All magnetic fields, the light vector, and sample surface normal are set to the horizontal plane. Note that the 3D vector magnet allows precise compensation of the residual fields. Magnetic fields parallel to the light wave vector \textbf{k} are denoted as $\textbf{B}_{\rm F}$ (Faraday geometry) and magnetic fields perpendicular to \textbf{k} as $\textbf{B}_{\rm V}$ (Voigt geometry). The angle $\theta$ defines the tilt of the magnetic field from the Faraday geometry.

\subsection{Photoluminescence}

The photoluminescence (PL) was excited by a continuous-wave laser with the photon energy of 3.06\,eV (405\,nm). The emitted light was coupled into a 0.5\,m monochromator equipped with a Peltier cooled charge coupled device (CCD) via a fiber. 

\subsection{Time-resolved Kerr ellipticity (TRKE)}

The coherent spin dynamics of electrons and holes interacting with the nuclear spins were measured by a degenerate pump-probe setup, where pump and probe have the same photon energy~\cite{yakovlevCh6}. A titanium-sapphire (Ti:Sa) laser generates 1.5\,ps long pulses with aspectral width of about $1$\,nm (about 1.5\,meV) and pulse repetition rate of 76\,MHz (repetition period $T_\text{R}=13.2$\,ns). The Ti:Sa laser beam was fed into an optical parametric oscillator with an internal frequency doubling and the output photon energy was adjusted to values around the exciton resonance to meet the maximum of the Kerr rotation signal at 2.191\,eV (566\,nm). The laser output was split into the pump and probe beams. The probe pulses were delayed relative to the pump pulses by a double-pass mechanical delay line with one meter length. The pump and probe beams were modulated using a photoelastic modulator (PEM) for the probe and an electrooptical modulator (EOM) for the pump. The probe beam was always linearly polarized and amplitude modulated at a frequency of  84\,kHz. The pump beam was either helicity modulated between the $\sigma^+/\sigma^-$ circular polarizations, or amplitude modulated with fixed helicity, either $\sigma^+$ or $\sigma^-$, in the frequency range from 0 to 5\,MHz. In all cases $f_{\rm mod}$ refers to the helicity modulation frequency. Amplitude modulation can be in effect considered as 0~Hz helicity modulation, as the signal is independent from the bare amplitude modulation frequency. In the experiment typically 20~Hz to 100~kHz helicity and amplitude modulation frequencies were used. The polarization of the reflected probe beam was analyzed in respect of the rotation of its elliptical polarization (Kerr ellipticity) with a balanced photodiode, using a lock-in technique.  

\section{Acknowledgments}

We thankful to  M. M. Glazov for fruitful discussions. This work was supported by the Deutsche Forschungsgemeinschaft through the International Collaborative Research Centre TRR 160 (Project A1). The work at ETH Z\"urich (O.H. and M.V.K.) was financially supported by the Swiss National Science Foundation (grant agreement 186406, funded in conjunction with SPP2196 through DFG-SNSF bilateral program) and by the ETH Z\"urich through the ETH+ Project SynMatLab. D.S.S. acknowledges the RF President Grant No. MK-5158.2021.1.2, and the Foundation for the Advancement of Theoretical Physics and Mathematics ``BASIS''. The theoretical modeling of DNSS formation by D.S.S. was supported by the Russian Science Foundation (grant No. 21-72-10035).

\let\addcontentsline\oldaddcontentsline% Restore \addcontentsline
\makeatletter
\renewcommand\tableofcontents{%
    \@starttoc{toc}%
}
\makeatother
\renewcommand{\i}{{\rm i}}

%\appendix

%%%%%%%%%%%%%%%%%%%%               SOM               %%%%%%%%%%%%%%%%%%%%%%%%%%%%%%%%%
\onecolumngrid
\vspace{\columnsep}
\begin{center}
\newpage
\makeatletter
{\large\bf{Supplementary Information\\``\@title''}}
\makeatother
\end{center}
\vspace{\columnsep}

% The Supplementary Material includes the following topics:

\twocolumngrid
% \hypersetup{linktoc=page}
% \tableofcontents
% \vspace{\columnsep}

\counterwithin{figure}{section}
\renewcommand{\thepage}{S\arabic{page}}
\renewcommand{\theequation}{S\arabic{equation}}
\renewcommand{\thefigure}{S\arabic{figure}}
\renewcommand{\bibnumfmt}[1]{[S#1]}
\renewcommand{\citenumfont}[1]{S#1}

\setcounter{page}{1}
\setcounter{section}{0}
\setcounter{equation}{0}
\setcounter{figure}{0}

\section{I. Experimental details}

\subsection{A. Photoluminescence and time-resolved Kerr ellipticity}

\begin{figure}[t]
	\includegraphics[width=1\linewidth]{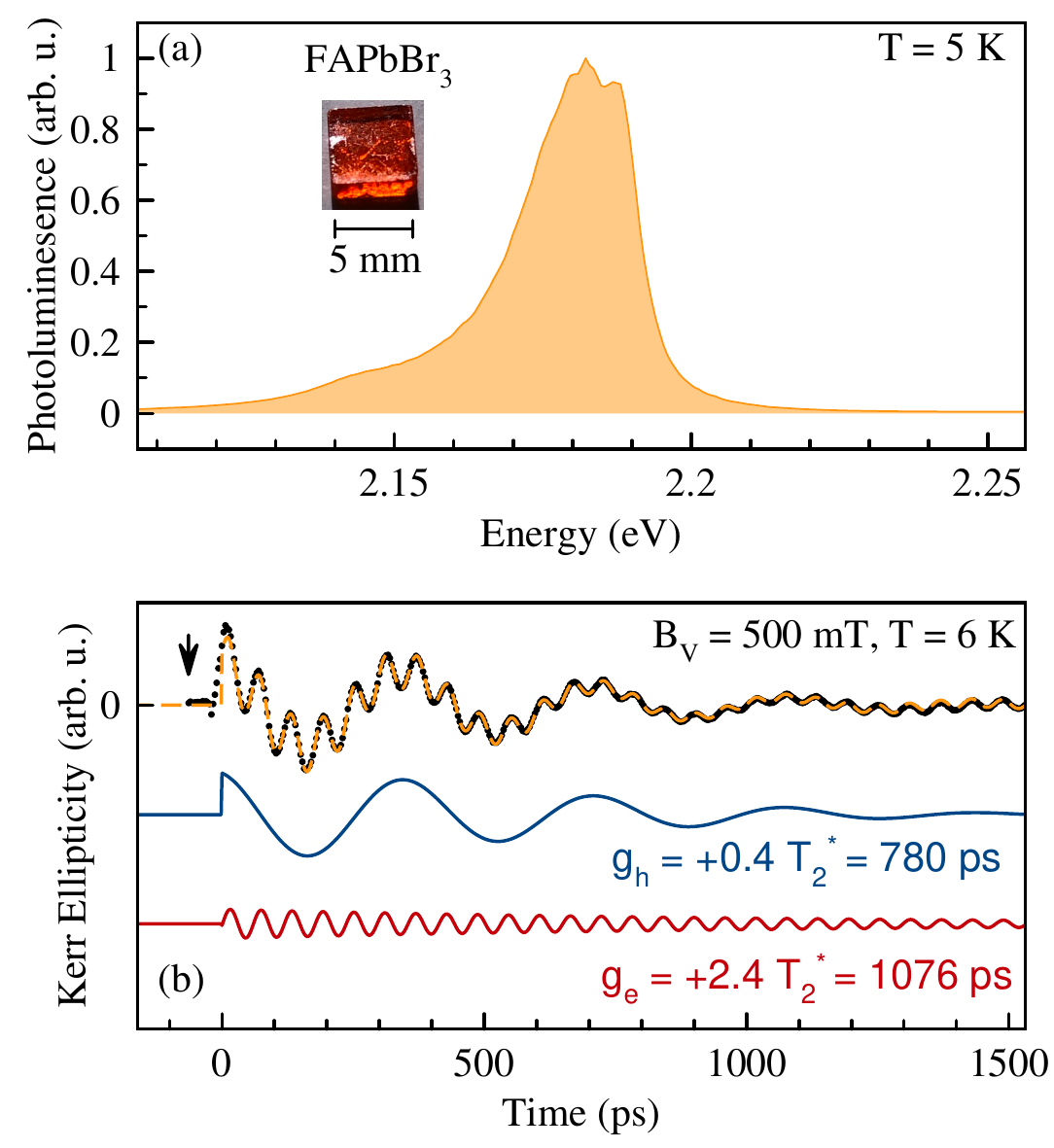}
        \caption{(a) Photoluminescence spectrum of the FAPbBr$_3$ crystal measured under continuous-wave excitation with the photon energy of 3.06~eV at $T=5$\,K. Inset shows a photo of the studied crystal. (b) Example of the TRKE dynamics (black line) at the pump-probe energy of 2.191\,eV. The pump beam was helicity modulated between $\sigma^+/\sigma^-$ circular polarization at the frequency of 50\,kHz. Its fit with Eq.~\eqref{eq:trke} is shown by the yellow dashed line. The electron and hole spin precession components evaluated from the fit are shown below by the red and blue lines, respectively.}
\label{fig:intro}
\end{figure}

A photo of the hybrid organic-inorganic FAPbBr$_3$ perovskite crystal under study is shown in the inset of Fig.~\ref{fig:intro}(a). It is a crystal with a size of 5 $\times$ 5 $\times$ 2~mm$^3$ and a glossy semi-transparent reddish color. Its low temperature photoluminescence measured at $T=5$\,K has maximum at 2.1776\,eV, see Fig.~\ref{fig:intro}(a). 

In the time-resolved Kerr ellipticity (TRKE) measurements the laser was tuned to the energy of 2.191\,eV, resonant with the free exciton. The pump beam was helicity modulated between $\sigma^+/\sigma^-$ circular polarization at the frequency of 50\,kHz to avoid dynamical nuclear polarization. A typical TRKE dynamics trace measured in $B_V=0.5$\,T magnetic field applied in the Voigt geometry is shown in Fig.~\ref{fig:intro}(b). It is contributed by the coherent spin dynamics of charge carriers, which spins are initially oriented by the circularly polarized pump and then precess about the magnetic field.  One can clearly see two Larmor precession frequencies, corresponding to electrons (e) and holes (h), which is the typical phenomenology for lead halide perovskite crystals~\cite{Kirstein_NC_2022}. We have shown that the coherent spin dynamics in perovskite semiconductors are provided by resident electrons and holes, which are localized in spatially different sites~\cite{S_belykh2019coherent,S_Kirstein2022ACS,S_kirstein2022}.

The TRKE dynamics were fitted using the function:
\begin{multline}
TRKE = \mathcal A_{e} \cos{(\Omega_{L,e} t)} \exp{(-t/T_{2,e}^*)}\\+ \mathcal A_{h} \cos{(\Omega_{L} t)} \exp{(-t/T_{2,h}^*)},
\label{eq:trke}
\end{multline}
as shown by the yellow dashed line in Fig.~\ref{fig:intro}(b). $\mathcal A_{e(h)}$ are the Kerr ellipticity amplitudes, reflecting the degrees of carrier spin polarization, $\Omega_{L}=g_{h}\mu_B B/\hbar$ and $\Omega_{L,e}=g_{e}\mu_B B/\hbar$ are the Larmor frequencies with the carrier $g$-factors $g_{e(h)}$, and $T_{2,e(h)}^*$ are the ensemble spin dephasing times. There we show the individual components of the hole (blue line) and the electron (red line) spin dynamics obtained form the fit with the parameters $g_h=+0.4$ and $T_{2,h}^*=780$\,ps for the holes and $g_e=+2.4$ and $T_{2,e}^*=1080$\,ps for the electrons. The assignment of the components is made based on the universal dependence of their $g$-factors on the band gap energy that we have published recently~\cite{Kirstein_NC_2022}.   

The time delay is fixed at the negative value of ${-75}$\,ps, marked by the arrow in Fig.~\ref{fig:intro}(b), for the measurements of the resonant spin amplification (RSA) and polarization recovery curves (PRCs).

\subsection{B. Nuclear spin polarization in tilted magnetic field}

\begin{figure*}[t]
   \includegraphics[width=0.99\linewidth]{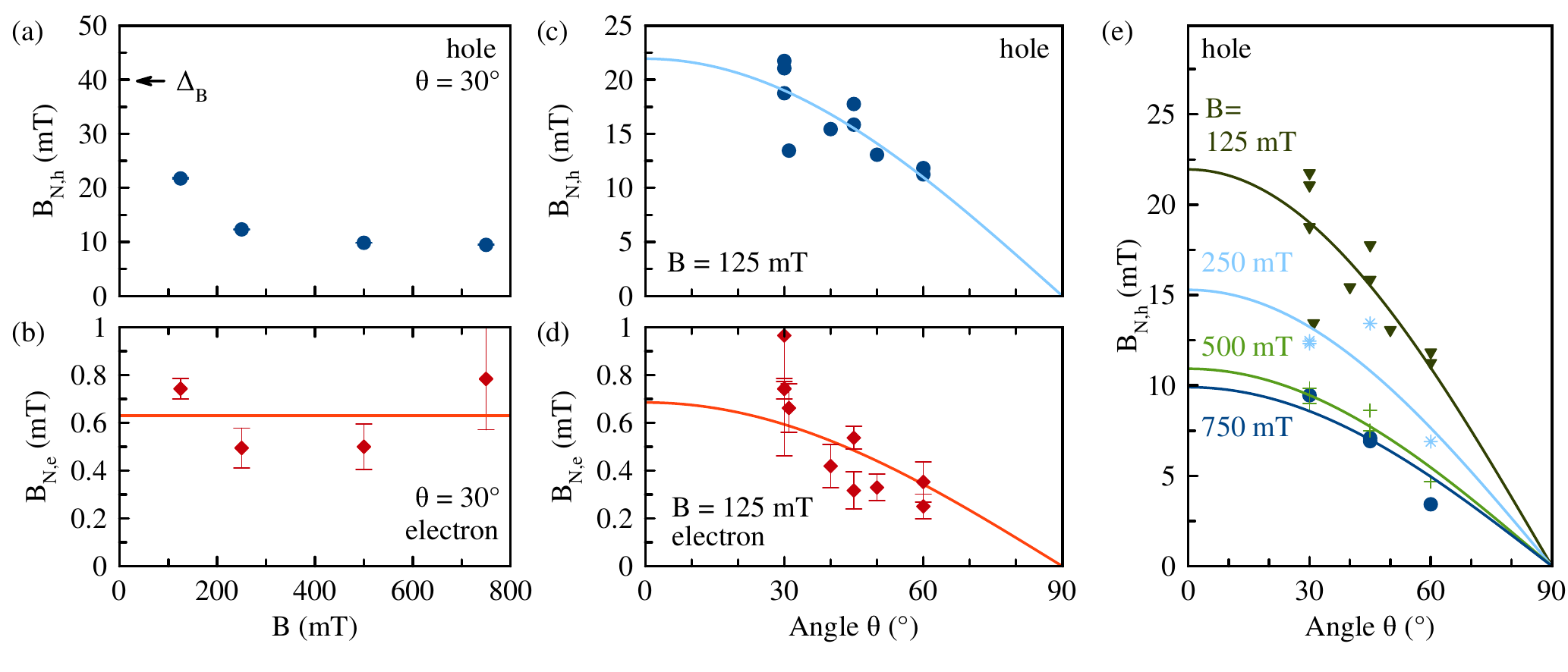}
 \caption{(a,b) Overhauser field ($B_{N,e(h)}$) experienced by the hole and electron spins as function of the magnetic field tilted by $\theta=30^\circ$. Lines are guides to the eye. (c,d) Overhauser fields as function of the tilting angle of the external magnetic field with strength $B=125\,$mT. Lines are fits according to $B_N \propto \cos{(\theta)}$. (e) Same as (c) for different magnetic field strengths. For all panels $T=1.6$\,K and the pump power is 8\,mW.}
 \label{fig:AngleDep}
 \end{figure*}

Using constant pump helicity, additional measurements of the nuclear spin polarization were performed using the same scheme as for Fig.~\ref{fig:dnptilted}. Figures~\ref{fig:AngleDep}(a,b) show that the Overhauser field for the electrons does not depend on the strength of the external magnetic field, as expected for dynamic nuclear spin polarization~\cite{S_book_Glazov}. The Overhauser field for the holes decreases from 22~mT at $B=150$~mT to 12~mT at $B=250$~mT and then stays about constant at 10~mT up to the magnetic field of 750~mT. Nevertheless, it remains small in comparison with its typical stochastic fluctuation in equilibrium, $\Delta_B=40$~mT. 

In the angular dependence of the Overhauser field for both electron and hole spins, despite a certain scattering of the achieved field value, a dependence according to a cosine-function is found, see Figs.~\ref{fig:AngleDep}(c,d). The angular dependence of the Overhauser field experienced by the holes for several magnetic fields is given in Fig.~\ref{fig:AngleDep}(e).

\begin{table}%
\caption{Isotopes with nonzero nuclear spin in FAPbBr$_3$ with FA = CH$_2$(NH)$_2$.}
\begin{tabular}{l|c|c|r}
Nuclear  & abundance & Spin & NMR $\gamma_N$ \\
isotope		& [\%] & $I$ & [MHz/T]  \\ \hline \hline
$^{207}$Pb & 22.10 & 1/2 & 8.882 \\
$^{79}$Br & 50.69 & 3/2 & 10.704 \\
$^{81}$Br & 49.31 & 3/2 & 11.538 \\
$^{1}$H & 99.98 & 1/2 & 42.577 \\ 
$^{13}$C & 1.07 &1/2 & 10.708 \\
$^{14}$N & 99.63 & 1& 3.078
\end{tabular}
\label{tab:nmr}
\end{table}

\subsection{C. PRC at constant helicity pumping}

 \begin{figure*}[t]
   \includegraphics[width=0.99\linewidth]{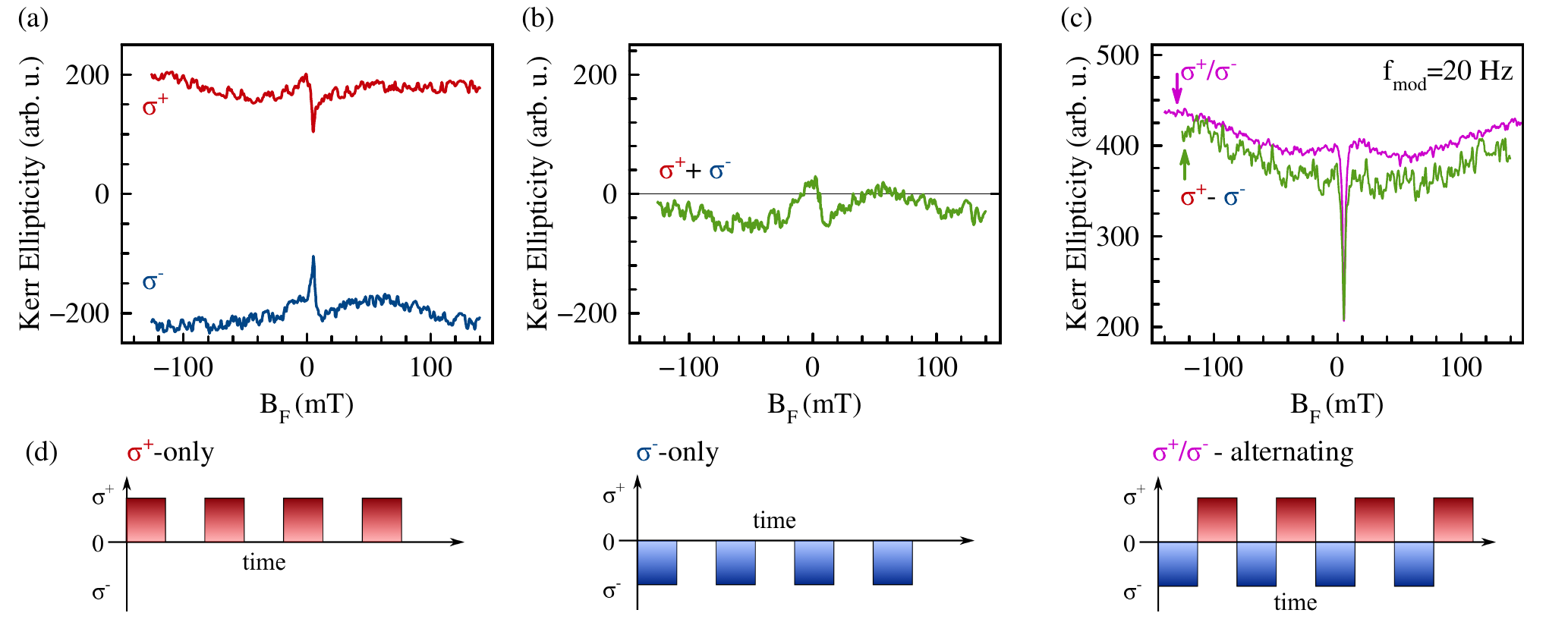}
   \caption{(a) PRC with constant pump helicity $\sigma^+$ (blue) or $\sigma^-$ (red). $T=1.6$\,K, pump power 6\,mW. (b) Sum of the two signals from panel (a). (c) Difference of the two signals from panel (a) (green), corresponding to the nuclear spin inertia signal at $f_{\text{mod}}=0$, and nuclear spin inertia signal at $f_{\text{mod}}=20$~Hz (magenta). (d) Scheme of the applied excitation.}
 \label{fig:PRCconst}
 \end{figure*}

In order to analyze the limit of $f_{\text{mod}}\to 0$, we performed additional PRC measurements for constant pump helicity, $\sigma^+$ or $\sigma^-$. Figure~\ref{fig:PRCconst}(a) shows the two corresponding Kerr ellipticity dynamics. They merge with each other when reversing amplitude and magnetic field. In particular, when their amplitudes are summed, the narrow peak around $B_F=0$ becomes fully compensated and cannot be seen [Fig.~\ref{fig:PRCconst}(b)].  We plot also the difference of the signals, which corresponds to the nuclear spin inertia signal at $f_{\text{mod}}=0$ and compare it with the case of $f_{\text{mod}}=20$~Hz in Fig.~\ref{fig:PRCconst}(c). Generally, the signals are similar, however the latter shows two weak maxima at $B_F\approx \pm 10$~mT around the zero-field dip in agreement with the theoretical modeling [Fig.~\ref{fig:teor}(f), green and blue curves].

\subsection{D. PRC power dependence}

 \begin{figure*}[t]
   \includegraphics[width=0.99\linewidth]{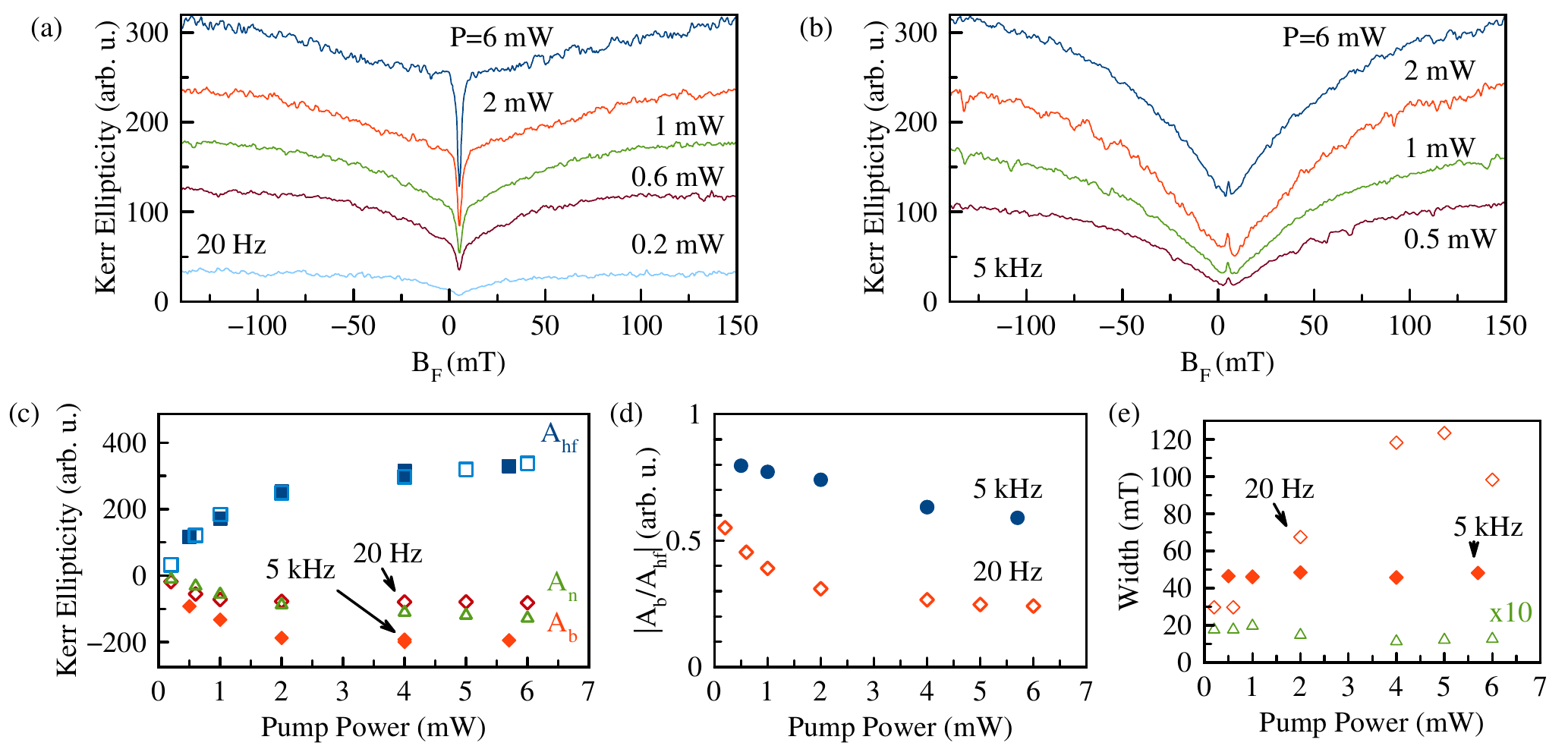}
   \caption{PRC at different pump powers $P$, for (a) $f_{\text{mod}}=20$\,Hz and (b) $f_{\text{mod}}=5$\,kHz. The curves are not shifted. (c,d,e)  Parameters evaluated from curves (a,b), full symbols for $f_{\text{mod}}=5$\,kHz and open symbols for $f_{\text{mod}}=20$\,Hz. (c) Pump power dependence of the amplitudes of the broad $A_b$ and narrow $A_n$ dips (red \& dark red diamonds and green triangles, respectively); high field (150~mT) values of the amplitudes $A_{\text{hf}}$ (blue \& light blue squares). (d) Pump power dependence of the ratio of amplitude of the broad dip and its high field value ($A_{\rm b}/A_{\rm hf}$) for $f_{\text{mod}}=20$\,Hz (red) and $f_{\text{mod}}=5$\,kHz (blue). (e) Pump power dependencies of the HWHM of the broad (open orange diamonds) and narrow (open green triangles) PRC dips at $f_{\text{mod}}=20$~Hz, and of the broad dip at $f_{\text{mod}}=5$~kHz (filled orange diamonds). For all panels $T=1.6\,$K.}
 \label{fig:PowerDep}
 \end{figure*}

Here we present data for the pump power dependence of the  PRC signals. In Figures~\ref{fig:PowerDep}(a,b) we compare the PRC curves for two different regimes: when the DNSS is formed ($f_{\rm mod}=20$\,Hz) and in absence of the DNSS ($f_{\rm mod}=5$\,kHz). 

All curves were evaluated by fits with respect to their amplitude, the ratio of broad dip ($A_{\rm b}$) to the high field ($A_{\rm hf}$) amplitude and the dip width, shown in Figs.~\ref{fig:PowerDep}(b,c,d), respectively. To start with 5\,kHz modulation frequency (full symbols), with increasing pump power the amplitude of the PRC dip and the high field amplitude rise linearly before showing a saturation trend. The narrow dip shows up as a minor peak with magnitude and width near the noise level so that is cannot be evaluated reliably for this frequency. The ratio of the PRC dip to the amplitude at high magnetic fields ($\approx 150\,$mT) [Fig.~\ref{fig:PowerDep}(c)] decreases slightly with increasing pump power. The width stays about constant in Fig.~\ref{fig:PowerDep}(d) for 5\,kHz modulation. 

In contrast, for the low modulation frequency of 20\,Hz, the PRC shape changes drastically with varying pump power. Again, the amplitudes at high field ($A_{\rm hf}$), of the broad dip ($A_{\rm b}$) and of the narrow dip ($A_{\rm n}$) rise nonlinearly with a tendency to saturate with increasing pump power. However, by contrast to the previous case, the ratio $A_{\rm b}$/$A_{\rm hf}$ here drops in a nonlinear fashion much stronger. Further, the width of the broad peak broadens significantly with excitation power. The width of the narrow peak remains constant. Note, in both cases of 5\,kHz and 20\,Hz, the power dependencies of the spin signals match each other at the high magnetic field of $B_F=\pm 150\,$mT, when the hyperfine interaction plays a minor role, see the blue symbols.

Overall the PRC pump power dependence for the low modulation frequency [Fig.~\ref{fig:PowerDep}(a)] is similar to the frequency dependence, shown in Fig.~\ref{fig:nucin}(b). As the DNSS formation rate is determined by the flux of angular momentum to the nuclear spin system, this shows that it is also proportional to the flux of angular momentum from exciting photons to a hole in agreement with the theoretical prediction, Eq.~\eqref{eq:nu0}.

\subsection{E. DNSS at zero magnetic field}
Surprisingly, we do not observe the hole spin polarization recovery with decreasing modulation frequency at zero magnetic field, which evidences the fragility of the DNSS in the field range $\lesssim1$\,mT. The exact reason for the DNSS not to form at zero field is unclear so far, but we note that the ODNMR in Fig.~\ref{fig:RSA} has the same linewidth of about $1$\,mT as the narrow PRC dip. This suggests that the DNSS is destroyed by the non-secular part of the nuclear dipole-dipole interaction. Application of the magnetic field with strengths exceeding that of the local fields suppresses this interaction and stabilizes the DNSS.

\section{II. Simulation of DNSS formation}
\label{sec:theory}

We describe the nuclear spin dynamics in the pump-probe experiments using the central spin box model. The Hamiltonian of the system reads
\begin{equation}
  \label{eq:Ham_SI}
  \mathcal H=A\bm I\bm S+\hbar\Omega_LS_z,
\end{equation}
where $\bm S$ is the hole spin, $A$ is the joint hyperfine interaction constant for all $N$ nuclei in the hole localization volume, $\bm I=\sum_{n=1}^N\bm I_n$ is the total nuclear spin composed of the individual spins $\bm I_n$. The hyperfine interaction is assumed here to be isotropic because of the dominant S-type Bloch wave functions at the top of the valence band in perovskites. The same description is valid for electrons in conventional GaAs-like semiconductors. However, in this work on perovskites the hyperfine interaction of the holes plays the main role.

\subsection{A. PRC without DNSS}

In equilibrium or at high enough polarization modulation frequencies, the nuclear spin distribution is Gaussian:
\begin{equation}
  \mathcal F(B_N)=\frac{\exp(-2B_N^2/\Delta_B^2)}{(\sqrt{\pi/2}\Delta_B)^{3}}
\end{equation}
with the parameter
\begin{equation}
  \label{eq:Delta_B}
  \Delta_B=\frac{\sqrt{N}A}{g_h\mu_B},
\end{equation}
determining the dispersion of the distribution. In these conditions, the PRC is described by~\cite{S_PRC}
\begin{equation}
  \text{TRKE}= 1-\frac{2}{3}\frac{\Delta_B^2}{\Delta_B^2+B^2}.
\end{equation}
One can see from this expression that the HWHM of the PRC is given by $\Delta_B$, which in combination with Eq.~\eqref{eq:Delta_B} yields
\begin{equation}
  N=(g_h\mu_B\Delta_B/A)^2,
\end{equation}
as used in the main text to determine the number of nuclei $N$.

\subsection{B. Nuclear spin dynamics}

To describe the nuclear spin dynamics, we note that in Eq.~\eqref{eq:Ham_SI} the total nuclear spin $I$ is conserved. For $N$ nuclear spins $1/2$, the number of possible realizations of spin $I$ is $C_N^{N/2-I}-C_N^{N/2-I-1}$, where $C_n^k$ is the binomial coefficient, and the probability to find the spin $I$ is
\begin{equation}
  \label{eq:P_I}
  P_I=\frac{(2I+1)(C_N^{N/2-I}-C_N^{N/2-I-1})}{2^N}.
\end{equation}
The eigenstates of the system can be labeled by the total angular momentum $F_z=I_z+S_z$, and have the form
\begin{equation}
  \Psi_\pm(F_z)=\mathcal A_\pm(F_z)\ket{F_z+1/2,\downarrow}+\mathcal B_\pm(F_z)\ket{F_z-1/2,\uparrow},
\end{equation}
where $\uparrow$ ($\downarrow$) corresponds to a hole with spin $S_z=+1/2$ ($-1/2$) and
\begin{subequations}
\begin{equation}
  \mathcal A_+(F_z)=-\mathcal B_-(F_z)=\frac{\Omega_x}{\sqrt{2\Omega(\Omega+\Omega_z)}},
\end{equation}
\begin{equation}
  \mathcal A_-(F_z)=\mathcal B_+(F_z)=\sqrt{\frac{\Omega+\Omega_z}{2\Omega}},
\end{equation}
\end{subequations}
with
\begin{subequations}
\begin{equation}
  \Omega_x=\frac{A}{\hbar}\sqrt{I(I+1)-F_z^2+1/4},
\end{equation}
\begin{equation}
  \Omega_y=0,
\end{equation}
\begin{equation}
  \Omega_z=\Omega_L+\frac{A}{\hbar}F_z
\end{equation}
\end{subequations}
and
\begin{equation}
  \Omega=|\bm\Omega|=\frac{1}{\hbar}\sqrt{A^2I(I+1)+\hbar^2\Omega_L^2+2AF_z\hbar\Omega_L+A^2/4}.
\end{equation}

For a given $I_z$ after initialization of the hole in the spin-up state, the wave function is $\ket{I_z,\uparrow}$, which can be represented as
\begin{equation}
  \ket{I_z,\uparrow}=\alpha\Psi_+(I_z+1/2)+\beta\Psi_-(I_z+1/2),
\end{equation}
where
\begin{subequations}
\begin{equation}
  \alpha^2=\frac{\mathcal A_-^2(I_z+1/2)}{\mathcal A_+^2(I_z+1/2)+\mathcal A_-^2(I_z+1/2)},
\end{equation}
\begin{equation}
  \beta^2=\frac{\mathcal A_+^2(I_z+1/2)}{\mathcal A_+^2(I_z+1/2)+\mathcal A_-^2(I_z+1/2)}.
\end{equation}
\end{subequations}
Then the coherence between the states $\Psi_+(I_z+1/2)$ and $\Psi_-(I_z+1/2)$ is lost~\cite{merkulov02} on the timescale $T_{2,h}^*\sim\hbar/(AI)\ll T_R$, where $T_R$ is the repetition period of the pump pulses. As a result, the nuclear spin component $I_z$ increases by unity with the probability
\begin{multline}
  v(I_z+1/2)=|\alpha\mathcal A_+(I_z+1/2)|^2+|\beta\mathcal A_-(I_z+1/2)|^2\\=\frac{2\mathcal A_-^2(I_z+1/2)\mathcal A_+^2(I_z+1/2)}{\mathcal A_-^2(I_z+1/2)+\mathcal A_+^2(I_z+1/2)}=\frac{1}{2}\left(\frac{\Omega_x}{\Omega}\right)^2.
\end{multline}
Here the argument $I_z+1/2$ is used to denote the total angular momentum component of the nuclei and the hole. In the limit of classical $I$, this reduces to $v(I_z+1/2)=A^2(I_x^2+I_y^2)/(2|A\bm I+\hbar\bm\Omega_L|^2)$. For a hole with spin down, the nuclear spin projection reduces by unity with the probability $v(I_z-1/2)$.

In the limit of many nuclear spins, $N\gg 1$, it is convenient to introduce the distribution function $g(x)$, where $x=I_z/I$, which is normalized such that
\begin{equation}
  \int\limits_{-1}^1g(x)\d x=1.
\end{equation}
It satisfies the Fokker-Planck equation
\begin{equation}
  I^2T_R\frac{\partial g}{\partial t}=P_e(t)I\frac{\partial}{\partial x}\left(vg\right)+\frac{\partial}{\partial x}\left(\frac{v}{2}\frac{\partial g}{\partial x}\right),
\end{equation}
where $v=(1-x^2)/2$. Introducing $\tau=t/(I^2T_R)$ and $\Pi(t)=P_e(t)I$, we obtain
\begin{equation}
  \label{eq:cont_dist}
  \frac{\partial g}{\partial\tau}=\Pi(t)\frac{\partial}{\partial x}\left(vg\right)+\frac{\partial}{\partial x}\left(\frac{v}{2}\frac{\partial g}{\partial x}\right).
\end{equation}
As boundary condition, $g(x)$ should be finite at $x=\pm1$.

\subsection{C. Nuclear spin inertia}

To describe the nuclear spin inertia, we assume that each pump pulse excites a trion with the probability $\Gamma_0$ and after its recombination, the hole spin polarization increases by $P_h$. According to the optical selection rules, the transverse hole spin components subsequently are erased~\cite{yugova09}. This allows us to neglect the off-diagonal components of the hole and nuclei density matrix. In this case, the nuclear spin state is described by the probabilities $P_m$ for finding the spin $I$ in the state with $I_z=m$. We note that after the hole spin decoherence, the hole spin reduces by a factor of $v(F_z)$, which provides an increase of $I_z$. Thus, if the polarization modulation period consists of $N_{\text{mod}}$ pulses, the signal measured by the nuclear spin inertia method is given by~\cite{PhysRevB.98.125306}
\begin{equation}
  L=\dfrac{\left|\sum_{k=1}^{N_{\text{mod}}/2}\sum_{m=-I}^IP_m(k)v(m+1/2)\right|}{\left|\sum_{k=1}^{N_{\text{mod}}/2}\exp\left(2\pi\i k/N_{\text{mod}}\right)\right|},
\end{equation}
where $P_m(k)$ describes the distribution of $I_z=m$ after $k$ pulses, starting from the beginning of the period, and we assume that it is sufficient to consider only the first half of the period.

This formalism allows us to simulate the evolution of the nuclear spin distribution function under pulsed excitation with modulation of the light helicity and to calculate the nuclear spin inertia signal as function of the magnetic field and the modulation frequency. Due to the numerical difficulties with the boundary conditions for Eq.~\eqref{eq:cont_dist}, we have used a finite number of nuclei, namely $N=1300$, as estimated in the main text.

To fit the frequency dependence of the PRC amplitude in Fig.~\ref{fig:teor}(d), we have used $\Gamma_0=0.03$ and $P_h=0.5$. The latter parameter determines the saturation of the PRC amplitude at low modulation frequencies, and the total DNSS formation rate
\begin{equation}
  \label{eq:nu0}
  \nu_0=\Gamma_0P_h/(T_RN)=0.9\,\text{ms}^{-1}
\end{equation}
determines the characteristic time scale of the nuclear spin dynamics. Thus, all parameters can be reliably determined from the experimental results.

\subsection{D. Nuclear spin squeezing and entanglement}

We characterize the shrinking of the nuclear spin distribution function along the transverse directions by the Kitagawa and Ueda spin squeezing parameter $\xi_s$~\cite{S_PhysRevA.47.5138,S_MA201189}, which is defined by
\begin{equation}
  \label{eq:xi_s_SI}
  \xi_s^2=\frac{4}{N}\braket{I_x^2}.
\end{equation}
During the simulation of the nuclear spin inertia, it can be extracted as
\begin{equation}
  \xi_s^2=\frac{4}{N}\left\langle\sum_{m=-I}^IP_m\frac{I(I+1)-m^2}{2}\right\rangle,
\end{equation}
where the angular brackets denote the averaging over the total nuclear spin $I$ with probabilities~\eqref{eq:P_I}, and $P_m$ symbolizes $P_m(k)$ with $k=1$, for which the spin squeezing reaches its maximum.

The result of the calculation of the nuclear spin squeezing degree is shown in Fig.~\ref{fig:teor}(e) by the solid line for the same parameters as in panel~(d). To find $\xi_s$ from the experimental results, we assume that the nuclear spin dynamics obey the evolution according to Eq.~\eqref{eq:Ham_SI}. Then, following the same procedure as described above, we find the DNSS that corresponds to the experimentally observed amplitude of the PRC at the given frequency and the calculated spin squeezing parameter for this state. The results are shown in Fig.~\ref{fig:teor}(e) by the dots.

For low modulation frequencies we obtain $\xi_s=0.29$, which is limited by nuclear spin diffusion due to an incomplete hole spin polarization. In the limit of a perfect nuclear spin alignment along the $z$ axis, the largest possible spin squeezing is $\xi_s=0.21$ because of the quantum fluctuations of the transverse components of the total nuclear spin $I\sim\sqrt{N}\sim36$. This situation can be called the Heisenberg limit, and it is only $1.4$ times smaller than the experimentally reached value.

The large nuclear spin squeezing suggests deep entanglement between nuclear spins. One of the strict evidences of entanglement is the violation of the generalized spin squeezing inequality \begin{equation}
  \label{eq:gss_SI}
  \braket{I_x^2}+\braket{I_y^2}+\braket{(I_z-\braket{I_z})^2}\ge\frac{N}{2}.
\end{equation}
We note, that this represents a particular case of Eq.~\eqref{eq:gss} with $M=N$, which means that there is at least any entanglement in the nuclear spins system.

The PRC directly measures the transverse nuclear spin components $\braket{I_x^2}+\braket{I_y^2}$. To check the violation of Eq.~\eqref{eq:gss_SI}, we consider the upper limit for the longitudinal nuclear spin fluctuations $\braket{(I_z-\braket{I_z})^2}\le N/4$, which corresponds to uncorrelated individual nuclear spins. From the definition in Eq.~\eqref{eq:xi_s_SI} we obtain that states with $\xi_s\le1/\sqrt{2}=0.707$ are entangled. This boundary is shown in Fig.~\ref{fig:teor}(e) by the blue dashed line.

For the lowest polarization modulation frequency of $f_{\text{mod}}=100$\,Hz, the PRC amplitude $[L(B_{\rm max})-L(B=0)]/L(B=0)=0.054$ corresponds to $I_x^2+I_y^2=0.054I^2$, which after averaging over $I$ yields $\braket{I_x^2+I_y^2}=0.04N$. Combined with the upper boundary for the longitudinal spin fluctuations we obtain 
\begin{equation}
  \braket{I_x^2}+\braket{I_y^2}+\braket{(I_z-\braket{I_z})^2}\le0.29 N,
\end{equation}
which violates Eq.~\eqref{eq:gss_SI} almost by a factor of two. In particular, using Eq.~\eqref{eq:gss} we find that the achieved DNSS is at least equivalent to an $M=0.58N=750$-body entanglement among the $N=1300$ nuclear spins.

This is a very conservative estimate mainly because of the assumption of unsuppressed longitudinal spin fluctuations. According to our model of DNSS formation, at the lowest polarization modulation frequency the DNSS steady state is reached, where the depth of entanglement is limited by the conservation of $I$ and by nuclear spin diffusion. For this state, the calculation of the nuclear spin fluctuations yields an $M=113$-body entanglement. We recall that the smaller $M$ is, the deeper is the entanglement.

%
%\bibliographystylesec{naturemag}
%\bibliographysec{Perovskite_FAPbBr_NMR}

%\bibliography{/home/dsmirnov/Practice/Theory/Lib/all-1}

%apsrev4-2.bst 2019-01-14 (MD) hand-edited version of apsrev4-1.bst
%Control: key (0)
%Control: author (8) initials jnrlst
%Control: editor formatted (1) identically to author
%Control: production of article title (0) allowed
%Control: page (0) single
%Control: year (1) truncated
%Control: production of eprint (0) enabled
%

\end{document}